\documentclass[12pt,oneside,reqno]{article}
\usepackage[utf8]{inputenc}
\usepackage{braket}
\usepackage{slashed}
\usepackage{graphicx}
\usepackage{comment}
\usepackage{amsmath}
\usepackage{amssymb}
\usepackage{amsthm}
\usepackage{amsfonts}
\usepackage{marginnote}
\usepackage{xcolor}
\usepackage{multicol}
\usepackage[margin=0.75in]{geometry}
\usepackage{hyperref}
\hypersetup{
    colorlinks=true,
    linkcolor=blue,
    filecolor=magenta,      
    urlcolor=cyan,
    citecolor=red,
    pdftitle={DiracFactorization},
    pdfpagemode=FullScreen,
    }
\graphicspath{ {./images/} }

\newtheorem{proposition}{Proposition}

\newcommand\blfootnote[1]{%
  \begingroup
  \renewcommand\thefootnote{}\footnote{#1}%
  \addtocounter{footnote}{-1}%
  \endgroup
}

\numberwithin{equation}{section}
\begin{document}

\title{Fractional Dirac Equations from Polynomial Linearization: Solutions and Difficulties}
\author{Erin T. Albertin$^{2}$, Zachary P. Bradshaw$^{1,*}$, Kaitlyn M. Kirt$^3$, \\
Kathryn E. Long$^3$, and Anthony Nguyen$^{2}$}
\date{
    $^1$Tulane University, Department of Mathematics, New Orleans, USA\\
    $^2$Tulane University, Department of Chemical and Biomolecular Engineering, New Orleans, USA\\
    $^3$Louisiana State University Math Circle, Baton Rouge, USA
}

\maketitle
\begin{abstract}
    The linearization of a quadratic form gives rise to a Clifford algebra structure, as seen in Dirac's factorization of the d'Alembert operator. A similar structure known as a generalized Clifford algebra arises from the continuation of this procedure to higher order forms. This technique combined with the existence of a fractional derivative satisfying the semi-group property can be used to factor the d'Alembert operator further, producing a fractional partial differential matrix equation that has a similar form to Dirac's equation. We examine these equations, their solutions, and point out difficulties when attempting to make physical sense of them.
\end{abstract}
\blfootnote{$*$ \textbf{Corresponding Author:} zbradshaw@tulane.edu}

\section{Introduction} \label{intro}
The linearization of a quadratic form is known to produce coefficients that form a representation of a Clifford algebra \cite{garling2011}, which is a quotient algebra of the tensor algebra by the ideal generated by elements of the form $v\otimes v-Q(v)1$ with $v$ an element of a quadratic space $(V,Q)$. The prime example is the Clifford algebra $Cl_{1,3}(\mathbb{C})$ generated by Dirac's $\gamma$-matrices obtained from the linearization of the d'Alembert operator $\Box=\Delta-\frac{1}{c^2}\partial_t^2$ \cite{dirac}, and the result of this procedure is Dirac's first order partial differential matrix equation 
\begin{align}\label{eq:dirac-equation}
    (i\hbar\gamma^\mu\partial_\mu-mc)\psi=0,
\end{align}
which he interpreted as a relativistic theory of the electron. Although, later this equation would be reinterpreted in the context of quantum field theory as a fundamental component of the standard model of particle physics \cite{Klauber,schwartz2014}.

When Dirac's procedure is extended to the linearization of higher order forms, the resulting coefficients form a representation of a generalized Clifford algebra wherein the quotient is taken with respect to the ideal generated by elements of the form $v^{\otimes k}-Q_k(v)1$, where $Q_k$ denotes a $k$-order form. This generalization was studied in detail in \cite{CHAPMAN2015,childs1978,micali2010,PAPPACENA2000}, and a procedure for the linearization of higher order forms is given in \cite{fleury1992}. Difficulties arise in the study of these generalized structures quickly. The Clifford algebra generated by a quadratic space $(V,Q)$, for example, has dimension $\dim(Cl(V,W))=2^{\dim(V)}$, while the dimension of a generalized Clifford algebra generated by a $k$-order form on a vector space $V$ is infinite whenever $\dim(V)>1$ and $k>2$ \cite{revoy2013}.

Dirac's factorization procedure can be extended using the linearization of higher order forms. In this setting, the d'Alembert operator is written as the $k$-th power of a linear polynomial of operators that correspond to fractional powers of the partial derivative operators (see \cite{raspini2000,raspini2001} for one take on the $k=3$ case). Such operators exist in the field of fractional calculus \cite{oldham}, which generalizes the standard derivative and integral operators to analogous operators of non-integer order. This field has developed rapidly in recent decades with the discovery of many applications \cite{hilfer}. The utility of most fractional derivatives in modeling physical systems arises from their non-local nature. Indeed, due to this non-locality, fractional derivatives have found use in modeling systems with memory \cite{herrmann,hilfer,oldham}. That is, systems in which the state at a time $t$ is dependent on the state of the system in a past interval of time.

In this work, we study the solutions and properties of the fractional Dirac equations that arise from this procedure. In Section~\ref{sec:poly-lin}, we review the procedure of Dirac and show how the coefficients in the linearization of a higher order form generate a representation of a generalized Clifford algebra. In Section~\ref{sec:apps}, we introduce a fractional derivative based on the Fourier transform to extend Dirac's factorization procedure. The solutions and properties of the resulting equation are then studied, and we discuss difficulties in interpreting this equation in any physically useful way. Finally, in Section~\ref{sec:conc}, we give concluding remarks.

\section{Polynomial Linearization} \label{sec:poly-lin}

Consider the quadratic form $p(x_1,\ldots,x_n)=x_1^2+\cdots+x_n^2$ lying in the polynomial ring $\mathbb{K}[x_1,\ldots,x_n]$ for some field $\mathbb{K}$. In some sense, what we want to do is take the square root of $p$; that is, we wish to find a linear polynomial $q(x_1,\ldots,x_n)$ such that $p(x_1,\ldots,x_n)=(q(x_1,\ldots,x_n))^2$. For $n=1$, the problem is trivial, but as soon as $n=2$, we find that if $q(x_1,x_2)=ax_1+bx_2$ exists, then the constraint
\begin{align*}
    (ax_1+bx_2)^2=a^2x_1^2+abx_1x_2+bax_1x_2+b^2x_2=x_1^2+x_2^2
\end{align*}
implies the conditions
\begin{align}
    a^2=b^2=1\label{eq:quadratic-condition1}\\
    ab+ba=0.\label{eq:quadratic-condition2}
\end{align}
Of course, if $a,b$ are elements of a field, this cannot be the case since the latter condition implies $a=0$ or $b=0$, so that the former condition is not satisfied. However, by replacing each $1\in\mathbb{K}$ by the identity $I\in M_2(\mathbb{K})$, $p$ can be promoted to an element of $M_2(\mathbb{K})[x_1,x_2]$, and because the matrix algebra $M_2(\mathbb{K})$ is non-commutative, the conditions \eqref{eq:quadratic-condition1} and \eqref{eq:quadratic-condition2} can now be satisfied. For example, the two Pauli matrices
\begin{align*}
    \sigma_x=\begin{pmatrix} 0&1\\
    1&0
    \end{pmatrix}\\
    \sigma_z=\begin{pmatrix}
        1&0\\0&-1
    \end{pmatrix}
\end{align*}
satisfy these conditions. Thus, $q(x_1,x_2)=\sigma_xx_1+\sigma_zx_2$ solves the linearization problem.

In general, one finds that the power-sum symmetric polynomial $p(x_1,\ldots,x_n)$ is linearized by a polynomial $q(x_1,\ldots,x_n)=\gamma^1x_1+\cdots+\gamma^nx_n$ with coefficients $\gamma^i\in M_m(\mathbb{K})$ for some $m$ if the following conditions are satisfied:
\begin{align*}
    \{\gamma^i,\gamma^j\}:=\gamma^i\gamma^j+\gamma^j\gamma^i=2\delta^{ij},
\end{align*}
where $i,j=1,\ldots,n$, $\delta^{ij}$ is the Kronecker delta function, and $\{a,b\}:=ab+ba$ denotes the anti-commutator. This more general case was considered by Brauer and Weyl \cite{brauer1935}, who showed that one solution to this problem is given by taking Kronecker products of the Pauli matrices and the two-dimensional identity matrix. Moreover, these coefficients form a representation of a Clifford algebra. To see this, let $V$ be a vector space with basis $\{e^1,\ldots,e^n\}$ and define a symmetric bilinear form $B(e^i,e^j)=\delta^{ij}$ with associated quadratic form $Q(e^i)=B(e^i,e^i)=1$. The Clifford algebra defined by a quadratic space is the quotient algebra $Cl(V,Q)=\mathcal{T}(V)/\langle v\otimes v-Q(v)1\rangle$ of the tensor algebra $\mathcal{T}(V)$ by the ideal generated by elements of the form $v\otimes v-Q(v)1$. Denote the product in the Clifford algebra by juxtaposition so that $v^2=Q(v)1$ for all $v\in V$. Then $(u+v)^2=Q(u)+Q(v)+\{u,v\}$ and $Q(u+v)=Q(u)+Q(v)+2B(u,v)$, from which it follows that $\{u,v\}=2B(u,v)$ for all $u,v\in V$. In particular, $(e^i)^2=1$ and $\{e^i,e^j\}=2\delta^{ij}$. By mapping $e^i\to\gamma^i$ and extending linearly, it follows that the coefficients $\gamma^i$ form a representation of this Clifford algebra. In the literature, another common notation for a Clifford algebra is $Cl_{p,q}(\mathbb{K})$ where $\mathbb{K}$ is the underlying field and there is a basis of $V$ such that $p$ basis vectors square to 1 and $q$ square to $-1$. The Clifford algebra we have constructed is therefore $Cl_{n,0}(\mathbb{K})$.

Consider the $k$-th order form given by the power-sum polynomial $p(x_1,\ldots,x_n)=x_1^k+\cdots+x_n^k$. A linearization $q(x_1,\ldots,x_n)=\gamma^1x_1+\cdots+\gamma^nx_n$ of $p$ has coefficients $\gamma^i$ satisfying
\begin{align*}
    \frac{1}{k!}\sum_{\sigma\in S_k}\gamma^{i_{\sigma(1)}}\cdots\gamma^{i_{\sigma(k)}}=\delta^{i_1\cdots i_k}
\end{align*}
where $S_k$ denotes the symmetric group on $k$ letters and $\delta^{i_1\cdots i_k}$ is a generalization of the Kronecker delta function which takes the value 1 when all indices are equal and 0 otherwise. Put another way, the symmetrization of a $k$-fold product of the coefficients vanishes unless the same coefficient is chosen for each index, in which case the symmetrization is unity. These coefficients have the structure of a generalized Clifford algebra, which is defined similarly to a Clifford algebra; given a $k$-order form $Q_k$, the associated generalized Clifford algebra is defined by $\mathcal{T}(V)/\langle v^{\otimes k}-Q_k(v)1\rangle$. Note how the Clifford algebra structure is recovered when $k=2$. In the power-sum polynomial example, a $k$-linear form $K$ is defined by $K(e^{i_1},\ldots,e^{i_k})=\delta^{i_1\cdots i_k}$ and the associated $k$-order form is $Q_k(v)=K(v,\ldots,v)$ so that, in particular, $Q_k(e^i)=1$. Denoting the product in the generalized Clifford algebra by juxtaposition, it follows that $v^k=Q_k(v)1$ for all $v\in V$; in particular, $(e^i)^k=1$. Since
\begin{align*}
    (e^1+\cdots+e^n)^k=\sum_{i_1,\ldots,i_k=1}^ne^{i_1}\cdots e^{i_k}
\end{align*}
and
\begin{align*}
    Q_k(e^1+\cdots+e^n)=\sum_{i_1,\ldots,i_k=1}^nK(e^{i_1},\ldots,e^{i_k}),
\end{align*}
it follows from a little induction that
\begin{align*}
    \frac{1}{k!}\sum_{\sigma\in S_k}e^{i_{\sigma(1)}}\cdots e^{i_{\sigma(k)}}=\delta^{i_1\cdots i_k}.
\end{align*}
By mapping $e^i\to\gamma^i$, it is now clear that the coefficients $\gamma^i$ form a representation of the generalized Clifford algebra given by the $k$-order form $Q_k$.

Note that this procedure can be extended even further by considering polynomial linearization with no assumption on the commutation relations between the linearization coefficients $\gamma^j$ and the variables $x_i$. It is not unreasonable, however, to make the assumption that $[x_i,x_j]=0$, as usual. Such is the case for the archetypal example of a partial differential equation, where the indeterminates $x_i$ are identified with the partial derivative operators $\partial_i$. The case of $[x_i,\gamma^j]=0$ for all $i,j$ is that of the generalized Clifford algebras which were studied in \cite{childs1978,morinaga1952}.

Consider again the power-sum symmetric polynomial $p(x_1,\ldots,x_k)=x_1^n+\cdots+x_k^n$. The case of $n=1$ is trivial. Let us consider instead the $n=2$ case. By invoking Einstein's summation convention, we write
\begin{align*}
    x_1^2+\cdots+x_k^2&=(\gamma^1x_1+\cdots+\gamma^kx_k)^2\\
    &=\gamma^ix_i\gamma^jx_j\\
    &=\gamma^i([x_i,\gamma^j]+\gamma^jx_i)x_j\\
    &=\gamma^i[x_i,\gamma^j]x_j+\gamma^i\gamma^jx_ix_j.
\end{align*}
If the commutation relations are independent of $x_i$ for all $i$, then we deduce the sufficient conditions
\begin{align}
    &\gamma^i[x_i,\gamma^j]=0,\label{eq:firstexp}\\
    &\gamma^i\gamma^j+\gamma^j\gamma^i=2\delta^{ij}\label{eq:secondexp},
\end{align}
where we recall that the index $i$ is summed over in \eqref{eq:firstexp}, and \eqref{eq:secondexp} is obtained from the assumption that $[x_i,x_j]=0$. 

Note how if $[x_i,\gamma^j]=0$, we are left with the defining relation for a Clifford algebra, which we denote by $Cl_{k,0}(\mathbb{C})$; however, in general the indeterminates $x_i$ are now coupled to the coefficients $\gamma^j$. Similarly, for the $n=3$ case, we have
\begin{align*}
    x_1^3+\cdots+x_k^3&=\gamma^{i_1}x_{i_1}\gamma^{i_2}x_{i_2}\gamma^{i_3}x_{i_3}\\
    &=(\gamma^{i_1}[x_{i_1},\gamma^{i_2}]x_{i_2}+\gamma^{i_1}\gamma^{i_2}x_{i_1}x_{i_2})\gamma^{i_3}x_{i_3}\\
    &=\gamma^{i_1}[x_{i_1},\gamma^{i_2}]x_{i_2}\gamma^{i_3}x_{i_3}+\gamma^{i_1}\gamma^{i_2}x_{i_1}x_{i_2}\gamma^{i_3}x_{i_3}\\
    &=\gamma^{i_1}[x_{i_1},\gamma^{i_2}][x_{i_2},\gamma^{i_3}]x_{i_3}+\gamma^{i_1}[x_{i_1},\gamma^{i_2}]\gamma^{i_3}x_{i_2}x_{i_3}\\
    &\ \ \ \ \ \ \ \ \ +\gamma^{i_1}\gamma^{i_2}x_{i_1}[x_{i_2},\gamma^{i_3}]x_{i_3}+\gamma^{i_1}\gamma^{i_2}x_{i_1}\gamma^{i_3}x_{i_2}x_{i_3}\nonumber\\
    &=\gamma^{i_1}[x_{i_1},\gamma^{i_2}][x_{i_2},\gamma^{i_3}]x_{i_3}+\gamma^{i_1}[x_{i_1},\gamma^{i_2}]\gamma^{i_3}x_{i_2}x_{i_3}\\
    &\ \ \ \ \ \ \ \ \ +\gamma^{i_1}\gamma^{i_2}[x_{i_1},[x_{i_2},\gamma^{i_3}]]x_{i_3}+\gamma^{i_1}\gamma^{i_2}[x_{i_2},\gamma^{i_3}]x_{i_1}x_{i_3}\nonumber\\
    &\ \ \ \ \ \ \ \ \ +\gamma^{i_1}\gamma^{i_2}[x_{i_1},\gamma^{i_3}]x_{i_2}x_{i_3}+\gamma^{i_1}\gamma^{i_2}\gamma^{i_3}x_{i_1}x_{i_2}x_{i_3}\nonumber\\
    &=(\gamma^{i_1}[x_{i_1},\gamma^{i_2}][x_{i_2},\gamma^{i_3}]+\gamma^{i_1}\gamma^{i_2}[x_{i_1},[x_{i_2},\gamma^{i_3}]])x_{i_3}\\
    &\ \ \ \ \ \ \ \ \ +(\gamma^{i_1}[x_{i_1},\gamma^{i_3}]\gamma^{i_2}+\gamma^{i_2}\gamma^{i_1}[x_{i_1},\gamma^{i_3}]+\gamma^{i_1}\gamma^{i_2}[x_{i_1},\gamma^{i_3}])x_{i_2}x_{i_3}\nonumber\\
    &\ \ \ \ \ \ \ \ \ +\gamma^{i_1}\gamma^{i_2}\gamma^{i_3}x_{i_1}x_{i_2}x_{i_3}\nonumber,
\end{align*}
from which we deduce the sufficient conditions
\begin{align}
    &\gamma^{i_1}[x_{i_1},\gamma^{i_2}][x_{i_2},\gamma^{i_3}]+\gamma^{i_1}\gamma^{i_2}[x_{i_1},[x_{i_2},\gamma^{i_3}]]=0\label{eq:cube1}\\
    &\sum_{\sigma\in S_2}\gamma^{i_3}[x_{i_3},\gamma^{i_{\sigma(1)}}]\gamma^{i_{\sigma(2)}}+\gamma^{i_{\sigma(2)}}\gamma^{i_3}[x_{i_3},\gamma^{i_{\sigma(1)}}]+\gamma^{i_3}\gamma^{i_{\sigma(2)}}[x_{i_3},\gamma^{i_{\sigma(1)}}]=0\label{eq:cube2}\\
    &\sum_{\sigma\in S_3}\gamma^{i_{\sigma(1)}}\gamma^{i_{\sigma(2)}}\gamma^{i_{\sigma(3)}}=6\delta^{i_1i_2i_3},\label{eq:cube3}
\end{align}
where again we notice that if $[x_i,\gamma^j]=0$ for all $i,j$, then we recover the generalized Clifford algebra structure discussed previously. The extension to power-sum polynomials with arbitrary coefficients is easily achieved. Letting
\begin{align*}
    p(x_1,\ldots,x_k)=a^ix_i^n,
\end{align*}
where we have again made use of Einstein's summation convention, we see that for $n=2$, the sufficient conditions \eqref{eq:firstexp} and \eqref{eq:secondexp} are modified as follows:
\begin{align}
    &\gamma^i[x_i,\gamma^j]=0,\label{eq:modifiedfirstexp}\\
    &\gamma^i\gamma^j+\gamma^j\gamma^i=2a^i\delta^{ij}\label{eq:modifiedsecondexp}.
\end{align}

It is perhaps worth illustrating this more general setting with an example. Consider the second order differential equation
\begin{align}
    x^2\frac{d^2}{dx^2}\psi(x)=c\psi(x),\label{eq:simpleexampleequation}
\end{align}
where $c$ is a constant. Letting $x_1=\frac{d}{dx}$ and $x_2=1$, we wish to write
\begin{align}
    x^2\frac{d^2}{dx^2}&=(\gamma^1(x)x_1+\gamma^2(x)x_2)^2\nonumber\\
    &=\left(\gamma^1(x)\frac{d}{dx}+\gamma^2(x)\right)^2. \label{eq:simpex}
\end{align}
Noting that $[\frac{d}{dx},\gamma^j(x)]=\frac{d\gamma^j(x)}{dx}$ and $[1,\gamma^j(x)]=0$, we may expand \eqref{eq:simpex}, from which we derive the sufficient conditions
\begin{align}
    &(\gamma^1)^2=x^2 \label{eq:condition1}\\
    &\gamma^1\frac{d\gamma^2}{dx}+(\gamma^2)^2=0\label{eq:condition2}\\
    &\gamma^1\frac{d\gamma^1}{dx}+\gamma^1\gamma^2+\gamma^2\gamma^1=0.\label{eq:condition3}
\end{align}
Note that we could have simply applied \eqref{eq:modifiedfirstexp} and \eqref{eq:modifiedsecondexp}; although, these conditions are clearly stronger than \eqref{eq:condition1}-\eqref{eq:condition3}, which we interpret as a manifestation of our ignorance in the analysis of the general case. With the help of \textsc{Mathematica}, a two dimensional representation for the coefficients $\gamma^j$ can be computed. One such representation is given by
\begin{align*}
    \gamma^1&=\begin{pmatrix}
    x&x\\0&-x
    \end{pmatrix}\\
    \gamma^2&=\begin{pmatrix}
    -1&-1\\
    1&1
    \end{pmatrix},
\end{align*}
and we note that if $\psi$ solves $(\gamma^1\partial_x+\gamma^2)\psi=\sqrt{c}\psi$, then it also solves \eqref{eq:simpleexampleequation}. Indeed, it can be verified that
\begin{align}
    \psi_0(x)=c\sqrt{x}\bigg(C_1\sqrt{x}^{\sqrt{4c+1}}+C_2\sqrt{x}^{\ -\sqrt{4c+1}}\bigg)\begin{pmatrix}
    1\\
    0
    \end{pmatrix}\label{eq:sol1}
\end{align}
and
\begin{align}
    \psi_1(x)=c\sqrt{x}\bigg(C_1\sqrt{x}^{\sqrt{4c+1}}+C_2\sqrt{x}^{\ -\sqrt{4c+1}}\bigg)\begin{pmatrix}
    0\\
    1
    \end{pmatrix}\label{eq:sol2}
\end{align}
satisfy this criteria for arbitrary constants $C_1$ and $C_2$.

It was shown in \cite{levy-leblond1967} that intrinsic spin is not, in fact, a manifestation of relativity, for it is produced by the linearization of the free Schr\"odinger equation. An application involving this line of reasoning to nuclear collective motion is outlined in \cite{greiner1988}. Likewise, it appears from \eqref{eq:sol1} and \eqref{eq:sol2} that a spin structure is again inherited from the linearization procedure in our example, lending credence to the idea that spin is not a manifestation of relativity. Furthermore, the factorization of partial differential equations is but one example of this more general notion of polynomial linearization. A nice discussion of this topic can be found in \cite{grigoryev2004}.

\section{Fractional Dirac Equation}\label{sec:apps}

A fractional derivative can be defined in a natural way using the Fourier transform. Indeed, let $\mathcal{F}$ denote the Fourier transform. We define the $\alpha$-order Fourier derivative by $\partial^\alpha(f)=\mathcal{F}^{-1}\{e^{\pi\alpha i/2}\omega^\alpha\mathcal{F}\{f\}\}$, so that when $\alpha$ is an integer, the ordinary integer derivatives are recovered. Moreover, this derivative satisfies the semi-group property $\partial^\alpha\partial^\beta=\partial^{\alpha+\beta}$. Just as Dirac factored the d'Alembertian into a linear operator, this semi-group property allows us to apply higher order polynomial linearization to factor the d'Alembertian further. Indeed, we identify the squares of the derivatives in each of the four variables with a power of the indeterminates of a polynomial: $\partial_i^2\longleftrightarrow x_i^k$ for $i=1,2,3$ and for the special case $i=0$, we make the identification $-\frac{1}{c^2}\partial_0^2\longleftrightarrow x_0^k$. We therefore expect that the coefficients in the linearization will form a representation of a generalized Clifford algebra.

Our aim is to find operators $A^0$, $A^1$, $A^2$, $A^3$ such that
\begin{align*}
    \bigg(A^1x_1+A^2x_2+A^3x_3+A^0x_0\bigg)^k
\end{align*}
reproduces the $k$-order power-sum polynomial in four variables $p(x_0,x_1,x_2,x_2)=x_0^2+x_1^2+x_2^2+x_3^2$. This constraint is satisfied if the following conditions hold:
\begin{align} \label{nthrelation}
    \frac{1}{k!}\sum_{\sigma\in S_k}A^{i_{\sigma(1)}}\cdots A^{i_{\sigma(k)}}=\delta^{i_1\cdots i_k},
\end{align}
where the sum is over all permutations $\sigma$ in the symmetric group $S_k$ and $\delta^{i_1\cdots i_k}$ takes the value 1 when all indices are equal and 0 otherwise. It is instructive to redefine the coefficients by $\beta^0:=A^0$ and $\beta^j:=e^{-i\pi/k}A^j$ for $j=1,2,3$, so that the defining relation \eqref{nthrelation} becomes
\begin{align}\label{eq:covariant-form-relation}
    \frac{1}{k!}\sum_{\sigma\in S_k}\beta^{\mu_{\sigma(1)}}\cdots\beta^{\mu_{\sigma(k)}}=g^{\mu_1\cdots\mu_k},
\end{align}
where
\begin{align*}
    g^{\mu_1\cdots\mu_k}:=\begin{cases} 1, & \mu_1=\cdots=\mu_k=0\\
    -1, & \mu_1=\cdots=\mu_k\ne0\\
    0, & \text{else}
    \end{cases}.
\end{align*}
The new coefficients $\beta^\mu$ form a representation of a generalized Clifford algebra which is isomorphic to the one represented by the original coefficients $A^i$. In Dirac's case, the coefficients $A^i$ represent the Clifford algebra $Cl_{4,0}(\mathbb{C})$, while his $\gamma$-matrices represent $Cl_{1,3}(\mathbb{C})$, and it follows that these algebras are isomorphic.

Following the approach of \cite{fleury1992}, we define matrices $P$, $J$, and $K$ component-wise by
\begin{align*}
    &P_{ij}=\delta_{i+1,j}\\
    &J_{ij}=\omega^j\delta_{ij}\\
    &K_{ij}=-\xi^{2j-1}\delta_{ij},
\end{align*}
where $\delta_{i+1,j}$ is defined to be $\delta_{i+1-k,j}$ whenever $i+1>k$, $\omega=e^{2\pi i/k}$ is a $k$-th root of unity, and $\xi=e^{\pi i/k}$. When $k$ is odd, we define the generalized Pauli matrices to be
\begin{align*}
    &\sigma^1=P\\
    &\sigma^2=JP\\
    &\sigma^3=J,
\end{align*}
and when $k$ is even, we set
\begin{align*}
    &\sigma^1=P\\
    &\sigma^2=KP\\
    &\sigma^3=J.
\end{align*}
We will make use of the following proposition later, and the proof is left to the appendix.
\begin{proposition}\label{prop:pauli-prop}
    The generalized Pauli matrices satisfy
    \begin{align*}
        (\sigma^j)^\dagger=(\sigma^j)^{k-1}
    \end{align*}
    for any choice of $k$.
\end{proposition}
The corresponding coefficients $\beta^\mu$ are given by $\beta^0=\sigma^3\otimes I$ and $\beta^j=\xi\sigma^2\otimes\sigma^j$ for $j=1,2,3$. Thus, the linearization we are looking for is $\beta^0x_1+e^{i\pi/k}\beta^1x_1+e^{i\pi/k}\beta^2x_2+e^{i\pi/k}\beta^3x_3$. Now replacing the indeterminates by their respective differential operators, we have derived the $2/k$-order operator equation
\begin{align*}
    \Box=(e^{i\pi/k}\beta^\mu\partial^{2/k}_\mu)^k.
\end{align*}
Moreover, substituting this into the Klein-Gordon equation
\begin{align*}
    \Box\psi=\frac{m^2c^2}{\hbar^2}\psi
\end{align*}
and taking the $k$-th root produces the fractional partial differential equation
\begin{align}\label{eq:fractional-dirac}
    (e^{i\pi/k}\hbar^{2/k}\beta^\mu\partial^{2/k}_\mu-(mc)^{2/k})\psi=0,
\end{align}
which we choose to call the fractional Dirac equation.

\subsection{Solutions}
Plane waves are a common guess for solutions to partial differential equations that arise in physical applications due to their behavior and the fact that they are eigenfunctions of the derivative operator. Similarly, they are eigenfunctions of the fractional derivative we have defined. Indeed, we have
\begin{align}\label{eq:fractional-action}
    \partial^\alpha_x(e^{ipx/\hbar})=\left(\frac{ip}{\hbar}\right)^\alpha e^{ ipx/\hbar}.
\end{align}

Letting $\psi=e^{ipx/\hbar}u_{\Vec{p}}$ where $u_{\Vec{p}}$ is a $k^2$-dimensional vector, the fractional Dirac equation \eqref{eq:fractional-dirac} reduces to
\begin{align}\label{reducedeq}
    (e^{2\pi i/k}(\beta^0 p_0^{2/k}+\beta^jp_j^{2/k})-m^{2/k}c^{2/k})u_{\Vec{p}}=0.
\end{align} 
Note that $u_{\vec{p}}$ is a nontrivial solution only if $\det(e^{2\pi i/k}\beta^\mu p_\mu^{2/3}-m^{2/3}c^{2/3})=0$, and evaluating this determinant yields the relativistic dispersion relation $E^2=(\Vec{p})^2c^2+m^2c^4$. Let us solve \eqref{eq:fractional-dirac} in the simple case that $\Vec{p}=0$, which corresponds to the rest frame. In this case, the fractional Dirac equation becomes
\begin{align*}
    (e^{2\pi i/k}\beta^0p_0^{2/k}-m^{2/k}c^{2/k})u_{\Vec{p}}=0.
\end{align*}
Using the relativistic dispersion relation and assuming $m\ne0$ produces
\begin{align} \label{eq:conditionequation}
    (e^{2\pi i/k}\beta^0-1)u_{\Vec{p}}=0.
\end{align}

Since $\beta^0=J\otimes I$, \eqref{eq:conditionequation} implies that the components of $u_{\Vec{p}}$ are zero except for the $k$ components which are free since $\omega^k-1=0$. Thus, there are $k$ linearly independent solutions of the form
\begin{align*}
    \psi_j=e^{ip\cdot x/\hbar}\Vec{e_j},
\end{align*}
where $\Vec{e_j}$ is the $j$-th standard basis vector in $\mathbb{C}^{k^2}$ and $j=k(k-2)+1,\ldots,k(k-1)$. Unsurprisingly, another set of solutions of the form $\psi=e^{-ipx/\hbar}$ can be derived similarly. Indeed, with the same assumptions, the fractional Dirac equation becomes
\begin{align*}
    (e^{4\pi i/k}\beta^0p_0^{2/k}-m^{2/k}c^{2/k})u_{\Vec{p}}=0,
\end{align*}
and applying the relativistic dispersion relation (with $m\ne0$) produces
\begin{align} \label{eq:conditionequation2}
    (e^{4\pi i/k}\beta^0-1)u_{\Vec{p}}=0.
\end{align}
Since $\beta^0=J\otimes I$, \eqref{eq:conditionequation2} implies that the components of $u_{\Vec{p}}$ are zero except for the $k$ components which are free since $\omega^k-1=0$. Thus, there are $k$ linearly independent solutions of the form
\begin{align*}
    \psi_j=e^{ip\cdot x/\hbar}\Vec{e_j},
\end{align*}
where $\Vec{e_j}$ is the $j$-th standard basis vector in $\mathbb{C}^{k^2}$ and $j=k(k-1)+1,\ldots,k^2$.

In Dirac's equation, there are plane wave solutions of both type $e^{ix\cdot p/\hbar}$ and type $e^{-ix\cdot p/\hbar}$ and the solution space is four-dimensional. The solutions we have derived mirror those arising from the free Dirac equation in the sense that they are plane waves multiplied by a vector that compensates for the matrix form of the equation. However, as $k$ increases, the dimension of the solution space increases quadratically. A possible way to interpret this increase is by the incorporation of higher order spin. Dirac's proposal was to interpret his equation as a relativistic theory of the electron, which is a spin-1/2 fermion. Similarly, we might think of the fractional Dirac equation as describing a particle of some different spin. Moreover, the sign of the exponent in the plane wave solution indicated whether the particle in question was a matter particle or an antimatter particle.

Notice that in the fractional solutions, all $2k$ linearly independent solutions are obtained using the last $2k$ entries of $u_{\Vec{p}}$. When the plus sign is chosen in the plane wave, the $k$ linearly independent solutions correspond to $k$ different spin states for a matter particle. When the minus sign is chosen, there are $k$ spin states for an anti-matter particle. Are there solutions which make use of the remaining $k(k-2)$ entries? These would correspond to another $k-2$ types of particles. Incredibly, the answer appears to be no. To see this, replace the exponential $e^{2\pi i/k}$ in \eqref{eq:conditionequation} with an arbitrary constant $C$ and notice that this forces the solution to be trivial unless $C$ happens to be a positive integer power of $\omega$, in which case $C^k-1=0$ and this frees $k$ variables. Now, $C$ comes from the fractional derivative as in \eqref{eq:fractional-action} and the factor of $e^{\pi i/k}$ already present in the fractional Dirac equation \eqref{eq:fractional-dirac}. The requirement that $C=\omega^m$ for some $m$ is equivalent to the condition $a^{2/k}e^{\pi i/k}=e^{2\pi mi/k}$ for some $a$, from which it follows that $a=e^{(2m-1)\pi i/2}$. That is, the only nontrivial solutions are given by $a=1,i,-1,-i$. The choices $a=i$ and $a=-i$ correspond to the solutions we have already given; however, the choices $a=1$ and $a=-1$ correspond to $\psi=e^{\pm p\cdot x/\hbar}u_{\Vec{p}}$, which are both unphysical, and mathematically ill-defined solutions (since the Fourier transform does not even converge in these cases). Thus, the continuation of Dirac's factorization procedure actually does not produce additional particle types.

\subsection{Difficulties}
Recall that it is possible to define an adjoint Dirac equation primarly due to the relationship $(\gamma^\mu)^\dagger=\gamma^0\gamma^\mu\gamma^0$ between the $\gamma$-matrices and their adjoints. Indeed, taking the adjoint of the Dirac equation yields
\begin{align*}
    0&=\psi^\dagger(-i\hbar(\gamma^\mu)^\dagger\partial_\mu-mc)\\
    &=\psi^\dagger(-i\hbar\gamma^0\gamma^\mu\gamma^0\partial_\mu-mc)\\
    &=\psi^\dagger\gamma^0(-i\hbar\gamma^\mu\gamma^0\partial_\mu-mc\gamma^0),
\end{align*}
and now multiplying on the right by $\gamma^0$ produces
\begin{align*}
    \psi^\dagger\gamma^0(-i\hbar\gamma^\mu\partial_\mu-mc)=0.
\end{align*}
By defining the Dirac adjoint as $\overline\psi:=\psi^\dagger\gamma^0$, we come to the adjoint Dirac equation:
\begin{align}\label{eq:adjoint-dirac}
    \overline\psi(-i\hbar\gamma^\mu\partial_\mu-mc)=0.
\end{align}
Note that here the partial derivative operator acts from the right on $\overline\psi$ so that what one really means by \eqref{eq:adjoint-dirac} in the standard left-action notation is
\begin{align*}
    -i\hbar\partial_\mu\overline\psi\gamma^\mu-mc\overline\psi=0.
\end{align*}
This allows for the computation of a conserved current $J^\mu$ in the following way. Multiply \eqref{eq:adjoint-dirac} from the right by $\psi$ and \eqref{eq:dirac-equation} from the left by $\overline\psi$. Now taking the difference produces
\begin{align*}
    0&=-i\hbar(\partial_\mu\overline\psi\gamma^\mu\psi+\overline\psi\gamma^\mu\partial_\mu\psi)\\
    &=-i\hbar\partial_\mu(\overline\psi\gamma^\mu\psi).
\end{align*}
Thus, $J^\mu=\overline\psi\gamma^\mu\psi$ is a conserved current. This is the conserved quantity given by Noether's theorem which corresponds to the global $U(1)$-symmetry in the Dirac equation. It is not clear how one would derive an analogous result in the fractional setting for two reasons. First, the nice relationship 
\begin{align}\label{eq:gamma-adjoint-relation}
    (\gamma^\mu)^\dagger=\gamma^0\gamma^\mu\gamma^0
\end{align} which allowed for the simple calculation involving the adjoint Dirac equation no longer holds. This identity follows from the hermiticity condition $(\gamma^0)^\dagger=\gamma^0$ and the anti-hermiticity conditions $(\gamma^j)^\dagger=-\gamma^j$ for $j=1,2,3$. A generalization with $(\beta^0)^\dagger=(\beta^0)^{k-1}$ and $(\beta^j)^\dagger=-(\beta^j)^{k-1}$ actually holds and follows directly from Proposition~\ref{prop:pauli-prop}, but this does not produce something as useful as \eqref{eq:gamma-adjoint-relation}. Note, however, that by multiplying by $\beta^0$ in the first relation and $\beta^j$ in the second and applying \eqref{eq:covariant-form-relation}, it follows that $\beta^\mu$ is unitary for all $\mu=0,1,2,3$.

The second reason it is not clear how to derive an analogous conserved quantity in the fractional setting is the lack of a fractional version of Noether's theorem. A first step would be to derive a fractional version of the Euler-Lagrange equation under the assumption that the Lagrangian of the system depends not on the derivative operator, but its fractional counterpart. To do this, we follow a procedure similar to that in \cite{agrawal2002}. Let $\mathcal{L}(\psi,\partial^\alpha\psi)$ be the Lagrangian of the system and let us assume that $\psi$ is sufficiently well-behaved so that the $\alpha$-order derivative exists. We wish to find the function $\psi^*$ among all functions satisfying the boundary conditions $\psi(a)=A$ and $\psi(b)=B$ for which the action $S[\psi]=\int_a^b\mathcal{L}(\psi,\partial^\alpha\psi)dx$ is at an extremum. Let $\epsilon\in\mathbb{R}$ and define a family of curves by $\psi(x)=\psi^*(x)+\epsilon\eta(x)$ satisfying the boundary condition (so that $\eta(a)=\eta(b)=0$). Taking the derivative with respect to $\epsilon$ produces
\begin{align*}
    \frac{dS}{d\epsilon}=\int_a^b\left(\frac{\partial\mathcal{L}}{\partial\psi}\eta+\frac{\partial\mathcal{L}}{\partial(\partial^\alpha\psi)}\partial^\alpha\eta\right)dx.
\end{align*}
Define the inner product $\langle f,g\rangle=\int_\mathbb{R} f^*(x)g(x)\ dx$. By Parseval's identity, we have (with $\alpha\ge0$)
\begin{align*}
    \langle \partial^\alpha f,g\rangle&=\langle \mathcal{F}^{-1}\{e^{\pi\alpha i/2}w^\alpha\mathcal{F}\{f\}\},g\rangle\\
    &=\langle e^{\pi\alpha i/2}w^\alpha\mathcal{F}\{f\},\mathcal{F}\{g\}\rangle\\
    &=\langle \mathcal{F}\{f\},e^{-\pi\alpha i/2}w^\alpha\mathcal{F}\{g\}\rangle\\
    &=e^{-\pi\alpha i}\langle f,\mathcal{F}^{-1}\{e^{\pi\alpha i/2}w^\alpha\mathcal{F}\{g\}\}\rangle\\
    &=\langle f, e^{-\pi\alpha i}\partial^\alpha g\rangle,
\end{align*}
from which it follows that
\begin{align*}
    \frac{dS}{d\epsilon}=\int_a^b\left(\frac{\partial\mathcal{L}}{\partial\psi}+e^{-\pi\alpha i}\partial^\alpha\frac{\partial\mathcal{L}}{\partial(\partial^\alpha\psi)}\right)\eta\ dx.
\end{align*}
Letting $\frac{dS}{d\epsilon}=0$ and noting that $\eta$ is arbitrary, we see that the action is extremal under the condition
\begin{align}\label{eq:fractional-Euler-Lagrange}
    \frac{\partial\mathcal{L}}{\partial\psi}+e^{-\pi\alpha i}\partial^\alpha\frac{\partial\mathcal{L}}{\partial(\partial^\alpha\psi)}=0,
\end{align}
and we call \eqref{eq:fractional-Euler-Lagrange} the fractional Euler-Lagrange equation. Note how the classical Euler-Lagrange equation is recovered with $\alpha=1$. To derive Noether's theorem, one makes use of the simple action of the derivative on a product:
\begin{align*}
    \frac{d}{dx}(f(x)g(x))=\frac{d}{dx}(f(x))g(x)+f(x)\frac{d}{dx}(g(x)).
\end{align*} 
Unfortunately, the action of our fractional derivative on a product is much more complicated; however, even if it behaved according to the same product rule, the theorem would connect symmetries not to conserved quantities, but to quantities that are unchanged by the action of a fractional derivative, and the physical meaning of this is unclear.

\section{Conclusions}\label{sec:conc}
While the fractional calculations in Section \ref{sec:apps} are interesting in their own right, it is not clear what use this formalism might have in quantum theory. In a future work, we will attempt to apply these results to the study of Feynman integrals --- which arise in precision calculations in high-energy particle physics --- by constructing a new regularization scheme and comparing it to existing methods. The idea is to factor the field equations defining quantum electrodynamics, for example, vary the fractional parameter, and hope that the natural loop integral divergences manifest themselves as poles in this parameter, which can then be removed via renormalization. 

Some work in the direction of fractional field theories has been done in \cite{herrmann2008} but using fractional derivatives that do not posses the semi-group property needed to perform Dirac's factorization procedure. In our future work, we will attempt to reproduce these results with the fractional derivative defined here. While these fractional field equations may or may not have physical significance, it may be possible to make use of them in a mathematically rigorous way through the Feynman integration means mentioned earlier or in some other way.

\section*{Declarations}
\subsection*{Funding} ZPB acknowledges funding from the Department of Defense's SMART scholarship program. This work was also partially supported by the Louisiana State University Math Circle.
\subsection*{Competing Interests} The authors declare no non-financial competing interests.
\bibliographystyle{plainurl}
\bibliography{Ref}

\begin{thebibliography}{10}

\bibitem{agrawal2002}
Om~P. Agrawal.
\newblock Formulation of euler–lagrange equations for fractional variational
  problems.
\newblock {\em Journal of Mathematical Analysis and Applications},
  272(1):368--379, 2002.
\newblock \href {https://doi.org/10.1016/S0022-247X(02)00180-4}
  {\path{doi:10.1016/S0022-247X(02)00180-4}}.

\bibitem{brauer1935}
Richard Brauer and Hermann Weyl.
\newblock Spinors in n dimensions.
\newblock {\em American Journal of Mathematics}, 57(2):425--449, 1935.

\bibitem{CHAPMAN2015}
Adam Chapman and Jung-Miao Kuo.
\newblock On the generalized clifford algebra of a monic polynomial.
\newblock {\em Linear Algebra and its Applications}, 471:184--202, 2015.
\newblock \href {https://doi.org/10.1016/j.laa.2014.12.030}
  {\path{doi:10.1016/j.laa.2014.12.030}}.

\bibitem{childs1978}
Lindsay~N. Childs.
\newblock Linearizing of n-ic forms and generalized clifford algebras.
\newblock {\em Linear and Multilinear Algebra}, 5(4):267--278, 1978.
\newblock \href {https://doi.org/10.1080/03081087808817206}
  {\path{doi:10.1080/03081087808817206}}.

\bibitem{dirac}
Paul Adrien~Maurice Dirac and Ralph~Howard Fowler.
\newblock The quantum theory of the electron.
\newblock {\em Proceedings of the Royal Society of London. Series A, Containing
  Papers of a Mathematical and Physical Character}, 117(778):610--624, 1928.
\newblock \href {https://doi.org/10.1098/rspa.1928.0023}
  {\path{doi:10.1098/rspa.1928.0023}}.

\bibitem{fleury1992}
N.~Fleury and M.~Rausch~de Traubenberg.
\newblock Linearization of polynomials.
\newblock {\em Journal of Mathematical Physics}, 33(10):3356--3366, 1992.
\newblock \href {https://doi.org/10.1063/1.529936}
  {\path{doi:10.1063/1.529936}}.

\bibitem{garling2011}
D.J.H. Garling.
\newblock {\em Clifford Algebras: An Introduction}.
\newblock London Mathematical Society Student Texts. Cambridge University
  Press, 2011.
\newblock URL: \url{https://books.google.com/books?id=RxNtliK3ROYC}.

\bibitem{greiner1988}
Martin Greiner, Werner Scheid, and Richard Herrmann.
\newblock Collective spin by linearization of the schr\"odinger equation for
  nuclear collective motion.
\newblock {\em Modern Physics Letters A}, 03(09):859--866, 1988.
\newblock \href {https://doi.org/10.1142/S0217732388001021}
  {\path{doi:10.1142/S0217732388001021}}.

\bibitem{grigoryev2004}
Dima Grigoryev and Fritz Schwarz.
\newblock Factoring and solving linear partial differential equations.
\newblock {\em Computing}, 73:179--197, 09 2004.

\bibitem{herrmann2008}
Richard Herrmann.
\newblock Gauge invariance in fractional field theories.
\newblock {\em Physics Letters A}, 372(34):5515--5522, 2008.
\newblock \href {https://doi.org/10.1016/j.physleta.2008.06.063}
  {\path{doi:10.1016/j.physleta.2008.06.063}}.

\bibitem{herrmann}
Richard Herrmann.
\newblock {\em Fractional {C}alculus}.
\newblock WORLD SCIENTIFIC, 2nd edition, 2014.
\newblock \href {https://doi.org/10.1142/8934} {\path{doi:10.1142/8934}}.

\bibitem{hilfer}
R~Hilfer.
\newblock {\em Applications of Fractional Calculus in Physics}.
\newblock WORLD SCIENTIFIC, 2000.
\newblock \href {https://doi.org/10.1142/3779} {\path{doi:10.1142/3779}}.

\bibitem{Klauber}
Robert~D. Klauber.
\newblock {\em {Student Friendly Quantum Field Theory:}: {Basic Principles and
  Quantum Electrodynamics}}.
\newblock Sandtrove Press, Fairfield, Iowa, 2013.

\bibitem{levy-leblond1967}
Jean-Marc Levy-Leblond.
\newblock Nonrelativistic particles and wave equations.
\newblock {\em Commun. Math. Phys.}, 6, 12 1967.
\newblock \href {https://doi.org/10.1007/BF01646020}
  {\path{doi:10.1007/BF01646020}}.

\bibitem{micali2010}
A.~Micali, R.~Boudet, and J.~Helmstetter.
\newblock {\em Clifford Algebras and their Applications in Mathematical
  Physics}.
\newblock Fundamental Theories of Physics. Springer Netherlands, 2010.
\newblock URL: \url{https://books.google.com/books?id=rRsMkgAACAAJ}.

\bibitem{morinaga1952}
Kakutarō Morinaga and Takayuki Nōno.
\newblock {On the Linearization of a Form of Higher Degree and its
  Representation}.
\newblock {\em Journal of Science of the Hiroshima University, Series A
  (Mathematics, Physics, Chemistry)}, 16(none):13 -- 41, 1952.
\newblock \href {https://doi.org/10.32917/hmj/1557367250}
  {\path{doi:10.32917/hmj/1557367250}}.

\bibitem{oldham}
K.B. Oldham and J.~Spanier.
\newblock {\em The Fractional Calculus: Theory and Applications of
  Differentiation and Integration to Arbitrary Order}.
\newblock Dover books on mathematics. Dover Publications, 2006.

\bibitem{PAPPACENA2000}
Christopher~J. Pappacena.
\newblock Matrix pencils and a generalized clifford algebra.
\newblock {\em Linear Algebra and its Applications}, 313(1):1--20, 2000.
\newblock \href {https://doi.org/10.1016/S0024-3795(00)00025-2}
  {\path{doi:10.1016/S0024-3795(00)00025-2}}.

\bibitem{raspini2000}
Andrea Raspini.
\newblock Dirac equation with fractional derivatives of order 2/3.
\newblock {\em Fizika B}, 01 2000.

\bibitem{raspini2001}
Andrea Raspini.
\newblock Simple solutions of the fractional dirac equation of order 2/3.
\newblock {\em Physica Scripta}, 64(1):20--22, jul 2001.
\newblock \href {https://doi.org/10.1238/physica.regular.064a00020}
  {\path{doi:10.1238/physica.regular.064a00020}}.

\bibitem{revoy2013}
Philippe Revoy.
\newblock Clifford algebras of forms of higher degrees.
\newblock {\em Advances in Applied Clifford Algebras}, 24:205 -- 212, 2013.

\bibitem{schwartz2014}
Matthew~D Schwartz.
\newblock {\em Quantum field theory and the standard model}.
\newblock Cambridge University Press, 2014.

\end{thebibliography}

\section*{Appendix}
The proof of Proposition~\ref{prop:pauli-prop} is as follows.
\begin{proof}
    \sloppy For the $j=1$ case, observe that $(\sigma^1)^\dagger_{ij}=\delta^{j+1,i}$, while $(\sigma^1)^{k-1}_{ij}=\delta_{i+1,\ell_1}\delta_{\ell_1+1,\ell_2},\cdots,\delta_{\ell_{k-2},j}=\delta_{i+k-1,j}=\delta_{i-1,j}=\delta_{i,j+1}$. For the $j=3$ case, observe that $(\sigma^3)^\dagger_{ij}=\omega^{-i}\delta_{ji}$, while $(\sigma^3)^{k-1}_{ij}=\omega^{\ell_1+\cdots+\ell_{k-2}+j}\delta_{i,\ell_1}\cdots\delta_{\ell_{k-2},j}=\omega^{(k-1)i}\delta_{ij}=\omega^{-i}\delta_{ji}$. The $j=2$ case is slightly trickier. Suppose $k$ is odd, so that $\sigma^2=JP$. Then the entries of $\sigma^2$ are given by $(\sigma^2)_{ij}=\omega^i\delta_{i+1,j}$. Now, on the one hand,
    \begin{align*}
        (\sigma^2)^\dagger_{ij}=\omega^{-j}\delta_{i,j+1}.
    \end{align*}
    On the other hand,
    \begin{align*}
        (\sigma^2)^{k-1}_{ij}&=\omega^i\delta_{i+1,\ell_1}\omega^{\ell_1}\delta_{\ell_1+1,\ell_2}\cdots\omega^{\ell_{k-2}}\delta_{\ell_{k-2}+1,j}\\
        &=\omega^{i+\ell_1+\cdots+\ell_{k-2}}\delta_{i+1,\ell_1}\cdots\delta_{\ell_{k-2}+1,j}\\
        &=\omega^{(k-1)i+(k-1)(k-2)/2}\delta_{i+k-1,j}\\
        &=\omega^{ki-i+k^2/2-3k/2+1}\delta_{i,j+1}.
    \end{align*}
    Observe that $\omega^{ki}=1$. Moreover, $\omega^{k^2/2}=-1$ since $k$ is odd. Similarly, $\omega^{-3k/2}=-1$. Putting this together produces
    \begin{align*}
        (\sigma^2)^{k-1}_{ij}&=\omega^{-i+1}\delta_{i,j+1}\\
        &=\omega^{-j}\delta^{i,j+1}\\
        &=(\sigma^2)^\dagger_{ij}.
    \end{align*}
    Finally, for the $j=2$ case with $k$ even, we have $\sigma^2=KP$, so that the entries of $\sigma^2$ are given by $(\sigma^2)_{ij}=-k^{2i-1}\delta_{i+1,j}$. Then on the one hand, we have
    \begin{align*}
        (\sigma^2)^\dagger_{ij}=-\xi^{1-2j}\delta_{i,j+1},
    \end{align*}
    while on the other hand,
    \begin{align*}
        (\sigma^2)^{k-1}_{ij}&=-\xi^{2i-1}\delta_{i+1,\ell_1}\xi^{2\ell_1-1}\delta_{\ell_1+1,\ell_2}\cdots\xi^{2\ell_{k-2}-1}\delta_{\ell_{k-2}+1,j}\\
        &=-\xi^{2(i+\ell_1+\cdots+\ell_{k-2})-k+1}\delta_{i+1,\ell_1}\delta_{\ell_1+2,\ell_2}\cdots\delta_{\ell_{k-2}+1,j}\\
        &=-\xi^{1-k+2(k-1)i+(k-2)(k-1)}\delta_{i+k-1,j}\\
        &=-\xi^{1-k+2(k-1)(j+1)+(k-2)(k-1)}\delta_{i,j+1}\\
        &=-\xi^{1-k+2kj+2k-2j-2+k^2-3k+2}\delta_{i,j+1}\\
        &=-\xi^{1-2k+2kj-2j+k^2}\delta_{i,j+1}.
    \end{align*}
    Since $k$ is even, $\xi^{k^2}=1$. Similarly, $\xi^{-2k}=\xi^{2kj}=1$, and it follows that
    \begin{align*}
        (\sigma^2)^{k-1}_{ij}=(\sigma^2)^\dagger_{ij}.
    \end{align*}
    This completes the proof.
\end{proof}
\end{document}